# Determination of the map of efficiency of the J-PET detector with the GATE package

P. Kowalski[1], L. Raczyński[1], T. Bednarski[2], P. Białas[2], E. Czerwiński[2], K. Giergiel[2],

Ł. Kapłon[2,3], A. Kochanowski[3], G. Korcyl[2], J. Kowal[2], T. Kozik[2], W. Krzemień[2],

M. Molenda[3], I. Moskal[2], P. Moskal[2], Sz. Niedźwiecki[2], M. Pałka[2],

M. Pawlik-Niedźwiecka[2], Z. Rudy[2], P. Salabura[2], N.G. Sharma[2], M. Silarski[2],

A. Słomski[2], J. Smyrski[2], A. Strzelecki[2], K. Szymański[2], W. Wiślicki[1], P. Witkowski[2],

M. Zieliński[2], N. Zoń[2]

[1]Świerk Computing Centre, National Centre for Nuclear Research, 05-400 Otwock-Świerk, Poland

[2]Faculty of Physics, Astronomy and Applied Computer Science, Jagiellonian University, 30-059 Cracow, Poland

[3]Faculty of Chemistry, Jagiellonian University, 30-060 Cracow, Poland

**Abstract:** A novel PET detector consisting of strips of polymer scintillators is being developed by the J-PET Collaboration. The map of efficiency and the map of geometrical acceptance of the 2-strip J-PET scanner are presented. Map of efficiency was determined using the Monte Carlo simulation software GATE based on GEANT4. Both maps were compared using method based on the $\chi^2$ test.

# Section 1: Introduction

The GEANT4 Application for Tomographic Emission (GATE) represents one of the most advanced specialized software packages for simulations [1] of the Positron Emission Tomography (PET) scanners [2]. GEANT4 is a toolkit for the simulation of the passage of particles through matter, using Monte Carlo methods [3]. Despite its complexity, GATE is easily configurable using script language.

In GATE, there are many tools for designing PET scanners: repeaters that allow to design periodic structures of scanners, possibility of usage of advanced four-dimensional phantoms or ability to simulate time-dependent phenomena (like breathing, or changing of source activity). Thanks to its simplicity and configurability, GATE package is used in many disciplines of medical physics to simulate complex devices or therapies. It can be also successfully used in the simulations of the Jagiellonian Positron Emission Tomograph (J-PET) device.

J-PET [4-6] device is a prototype PET scanner, that uses plastic scintillators. Its main advantage, in comparison to known solutions, is the possibility of scanning 3D region of the patient, not just 2D slice. It will be also much cheaper than existing scanners.

The prototype scanner is planned to be built of detectors, that are placed on the lateral area of the cylinder with the diameter of 70 cm. The axis of each detector is parallel to the axis of the cylinder. Each detector is built of one strip of plastic scintillator and two photomultipliers attached to its ends (more details about the structure of the J-PET device in ref. [4]).

The aim of this work was to determine the two-dimensional map of efficiency of the 2-strip J-PET system and to prepare a tool for computing the map of efficiency for systems built of more than two detectors. However, it is important to stress, that due to the axial symmetry of the J-PET scanner, a 2-strip module should reflect main features of the full detector setup.

The map of efficiency is important for determining the maximum spatial resolution that could be achieved in the J-PET scanner. It will also help in understanding some effects connected with physical limitations of the spatial resolution in the J-PET scanner. In general the efficiency map is instrumental for optimization of the detector's performance.

**1.1. Map of efficiency**

Let us assume, that points of annihilation (points of generation of back-to-back gamma quanta) are generated uniformly in the region of the cylinder with length $Z_0$ and radius $R_0$. This cylinder represents the inside of the PET scanner. Inner space of this cylinder is discretized virtually into voxels with size dx x dy x dz (where dx=dy=dz=0.5 cm and dx x dy means the fragment of the cross section of the cylinder). The 3D map of efficiency is defined as:

$$f_{eff}(x,y,z) = \frac{N_{det}(x,y,z)}{N_{gen}(x,y,z)} \quad (1)$$

where (x,y,z) are the spatial coordinates of the center of the voxel, $N_{gen}$ is the number of all annihilations generated in the voxel and $N_{det}$ is the number of annihilations that could be classified as events (scintillations were detected in two detectors – voltage signals occurred in four photomultipliers attached to two different scintillators).

For the 2-strip J-PET system (Fig. 1), we simplify above definition. Points of annihilation must be generated only in the rectangular region between two walls of both scintillators. Annihilation generated outside this region cannot be detected. There may be also some angular cuts for the directions of generated back-to-back gamma quanta. These directions should ensure that at least one of quanta will hit one of scintillators.

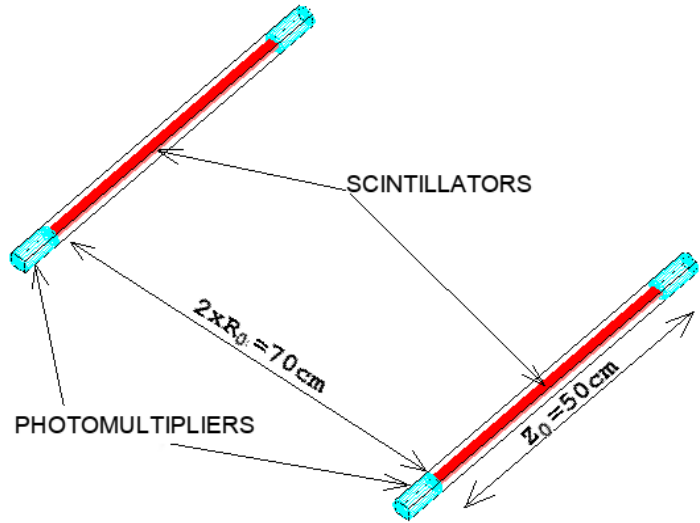

*Fig. 1. Visualization of 2-strip J-PET system*

The 3D voxels may be replaced with 2D pixels (with size dx x dz) of the plane which passes through both scintillators.

**1.2. Geometrical acceptance**

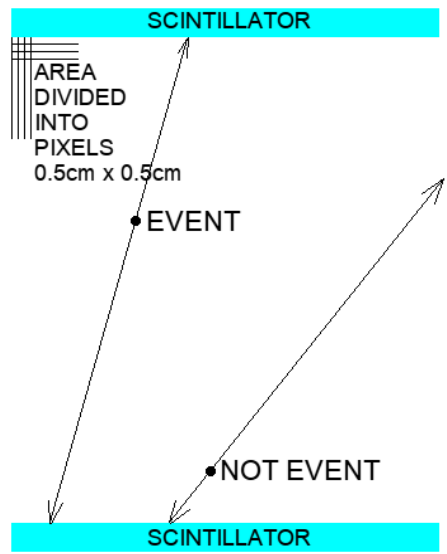

*Fig. 2. Method of estimating geometrical acceptance*

Geometrical acceptance does not include any physical effects, like interaction of gamma quanta with material of scintillators. The 2D map of geometrical acceptance of the 2-strip J-PET scanner was simulated independently of the GATE software using dedicated program based on the Monte Carlo method and written by the authors.

Inside the rectangular region with size 70 cm x 50 cm, point of annihilation was chosen randomly. After that the direction of back-to-back gamma quanta was randomized. If line with chosen direction crossing the point of annihilation was also crossing both shorter edges of the rectangle (corresponding to the scintillators), this case was treated as an event and was added to a histogram. The method of calculating geometrical acceptance is illustrated in Fig. 2.

The 2D map of geometrical acceptance may be defined as:

$$f_{acc}(x,z) = \frac{N_{acc}(x,z)}{N_{gen}(x,z)} \quad (2)$$

where (x,z) are the spatial coordinates of the center of the pixel (y=0), $N_{gen}$ is the number of all annihilations generated in the pixel and $N_{acc}$ is the number of all events caused by annihilations from the pixel.

## Section 2: Description of the simulation setup in the GATE software

In simulations, substance called EJ230 was used as a material of scintillators. Approximated chemical composition of this hydrocarbon polymer is $C_{10}H_{11}$. Some of its features are presented in Tab. 1.

| Feature | Value |
|---|---|
| State | solid |
| Density | 1.023 g/cm3 |
| Scintillation Yield | 10240 1/MeV |
| Refractive Index | 1.58 |
| Absorption Length | 110 cm |

*Tab. 1. Features of EJ230 material [7]*

Each scintillator is a cuboid with length equal to 50 cm, and rectangular cross-section 5 mm x 19 mm. Centre of the scintillator was placed in the zero of the x-axis and the scintillator was parallel to the z-axis of the coordinate system (Fig. 3).

In the simulation, the photomultipliers are modeled as a dielectric-metal interfaces. All optical photons hitting these surfaces are assumed to be detected. The distance between two detectors (each detector consists of one scintillator and two photomultipliers) was 70 cm.

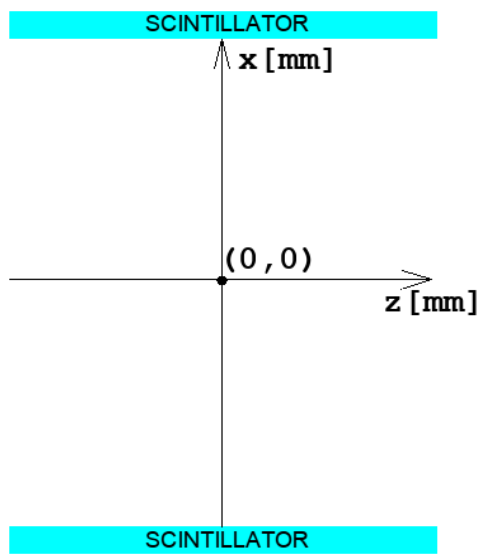

*Fig. 3. System of the 2-detector scanner in GATE coordinate system*

Source of back-to-back gamma quanta was a rectangular region stretched between two scintillators (walls 5 mm x 50 cm). Its size was 70 cm x 50 cm x 5 mm, and gamma quanta had energy of 511 keV (source illustrated in Fig. 4). In order to speed up the simulation, some limitations on the direction of generated gamma quanta were used.

The following physical processes were simulated: Compton effect, electron ionization, multiple electron scattering, fluorescence, optical absorption, scintillation and boundary effects.

The simulation was performed using computing cluster at Świerk Computing Centre Project (CIŚ) at National Centre for Nuclear Research.

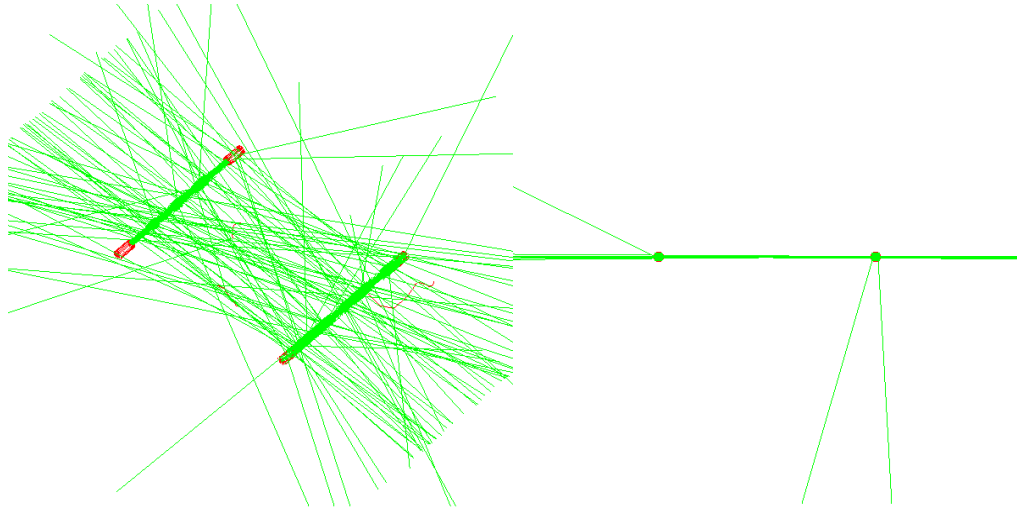

*Fig. 4. Visualization of the simulation of 2-strip J-PET system by means of the GATE software; lines represent back-to-back gamma quanta. One can see that some of gamma quanta underwent scattering in the detector. Left panel shows the perspective view of the 2-strip system and right panel shows the cross section in the plane perpendicular to the tomograph axis.*

Thanks to the analysis of the ROOT [8] output, obtained with GATE software (version 6.2), the map of efficiency was calculated using specialized tools developed in *python* and *C++*. Furthermore, some additional properties of the studied system could be checked, like wavelengths spectrum of photons that were detected by dielectric-metal

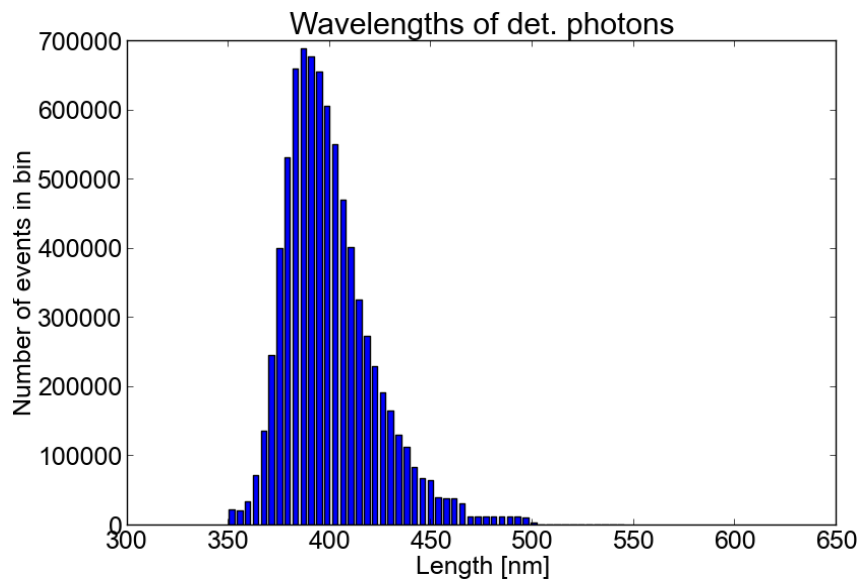

*Fig. 5. Distribution of wavelengths of photons detected by dielectric-metal surfaces (photomultipliers) in GATE software – histogram of 8 mln of photons; wavelengths accepted by photomultipliers R4998 are in range between 300 nm and 650 nm*

surfaces (Fig. 5). This exercise showed, that wavelengths of generated photons are consistent with wavelength accepted by photomultipliers used in the prototype J-PET device (Hammamatsu R4998 [9]). Furthermore, emission spectrum of scintillator simulated in GATE software is in accordance with spectrum obtained in the experiment.

## Section 3: Results and discussion

Map of geometrical acceptance was calculated using 100 mln generated points of annihilation. The rectangle was divided into 14000 bins with size 0.5 cm x 0.5 cm corresponding to about $7 \cdot 10^3$ points generated in each bin. Normalized (by the maximum value) map of geometrical acceptance is presented in Fig. 6.

The map of efficiency was calculated using much poorer statistics. Modesty of the event sample was due to the long simulation time. Presented map of efficiency (Fig. 7) consists of about 210000 points of annihilation.

All annihilations that caused coincidences (scintillations occurred in both scintillators at the same time), were treated as events. Both processes: interaction of gamma quantum in the scintillator material and the transport of photons inside the scintillator, were simulated. However, a number of photons produced in the process of the scintillation was not taken into account. This means that some of annihilations counted as an event, in real world would not be detected because of too small number of photons reaching photomultipliers (signals at photomultipliers could have too small amplitudes). If there would be an amplitude filter, the event sample would be smaller.

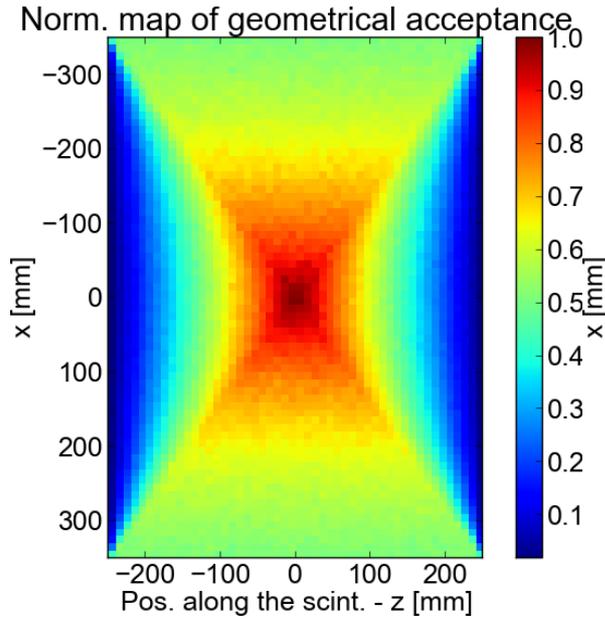 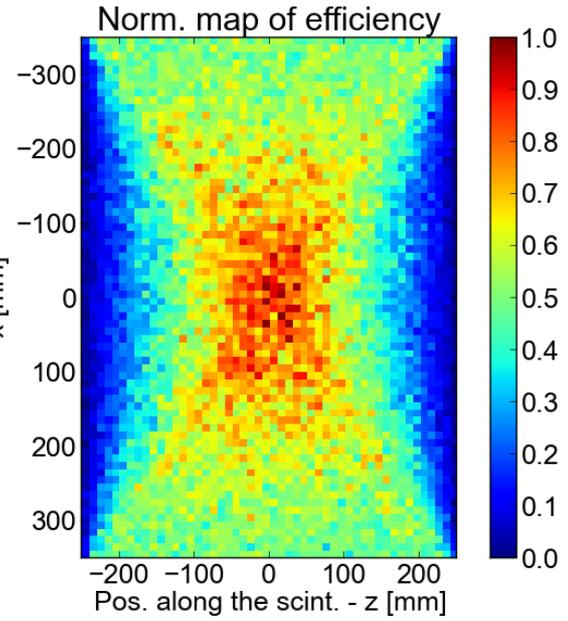

*Fig. 6. Map of geometrical acceptance*  *Fig. 7. Map of efficiency calculated using GATE software*

The map of efficiency (normalized by the maximum value), computed using GATE software and dedicated tools, is presented in Fig. 7. Maximum number of events per bin was about 120.

### 3.1. Comparison – the $\chi^2$ test

The corresponding values of pixels of two maps (Fig. 6, Fig. 7) were treated as random variables with the Poisson distribution. After normalization of the map of geometrical acceptance, it was possible to show what were the statistical dependencies between both maps. In calculations, it was assumed that if number of counts was greater than 10 [10], Poisson distribution could be approximated with the normal distribution.

Each pair of pixels was analyzed independently (with assumption that the values of pixels in each map were independent) and their values were compared using the $\chi^2$ test (Eq. 3). The value of $\chi^2$ test between two pixels located in the same position (i,j) in two maps (x and y), can be defined as:

$$\forall_{i,j} \chi_{ij}^2 = \frac{e_{ij}^2}{E_{ij}} \quad (3)$$

where $e_{i,j} = x_{i,j} - \frac{y_{i,j}}{k}$, $E_{i,j} = x_{i,j} + \frac{y_{i,j}}{k^2}$ and $k = \frac{\sum x_{i,j}}{\sum y_{i,j}}$. The calculated map of the value of $\chi^2$ test is presented in Fig. 8.

Values of the $\chi^2$ test were calculated for pixels, where number of counts in corresponding pixels of the map of efficiency (from GATE software), were greater than 10. In the opposite situation, as it is seen in side areas of the picture, values of the $\chi^2$ test were not calculated – white colour.

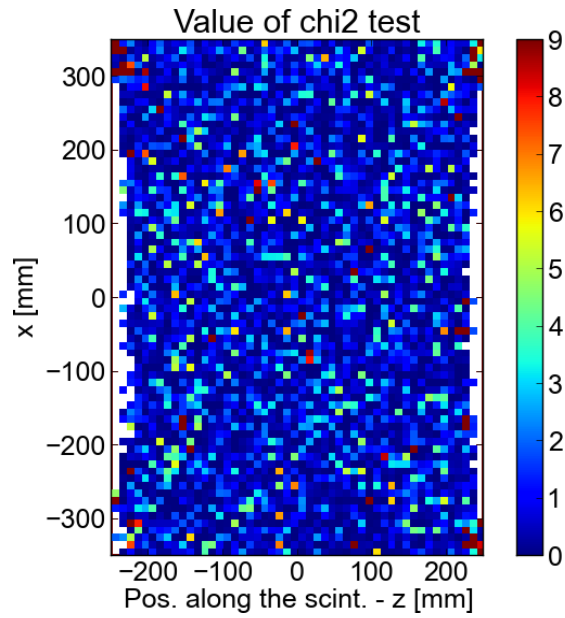

*Fig. 8. The map of the value of $\chi^2$ test*

### 3.2. Detector effects

In order to separate detector effects from geometrical acceptance, the map of efficiency $f_{eff}$ was divided (pixel by pixel) by the map of geometrical acceptance $f_{acc}$. Before calculating new map, the map of geometrical acceptance was normalized (divided) by the factor k:

$$k = \frac{S_x}{S_y} \quad (4)$$

where $S_x$ is number of all events in the map of geometrical acceptance and $S_y$ is number of events in the map of efficiency. Calculated map is presented in Fig. 9.

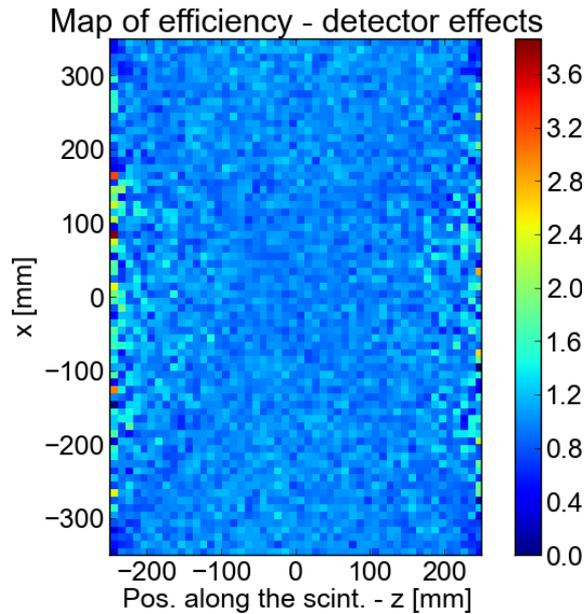
*Fig. 9. Map of efficiency - only detector effects are included*

As one can see, most of the central area of the map showing detector effects, is homogeneous and has value about 1. It means that influence of detector effects on the map of efficiency is much less than influence of geometrical acceptance. In boundary regions of the scanner, values of pixels from map of efficiency are even 3 times bigger than corresponding values of the map of geometrical acceptance. It may mean that in these regions, physical phenomena has significant influence on efficiency of the scanner. When other physical effects, like amplitudes of voltage signals, will be included in analysis, this effect will be studied more accurately.

It may be also observed, that on the diagonals, the efficiency is less than in other regions. This effect may be caused by the fact, that gamma quanta generated in these areas with maximum angle which may lead to coincidence, have smaller paths in scintillators and it is less probable that they will cause scintillations.

**3.3. Discussion of results**

At first sight, the map of the geometrical acceptance and the efficiency calculated using GATE look very similar. Chosen method of comparison showed that there are no statistically significant differences between two compared maps (Fig. 6, Fig. 7). For the majority of pixels, the $\chi^2$ values are much less than 9, which corresponds to three-sigma interval. It is worth to note, that there is no structure visible in Fig. 8. It shows that the map of efficiency is strongly dominated by geometrical properties of the system and detector effects are not clearly visible. However, the map of efficiency including only detector effects, showed that these effects influence on final efficiency of the scanner.

Results of performed calculations will be the subject of further studies. The map of efficiency with much greater statistics will be computed and taken to calculations of presented comparison map. Also other methods of comparison, e.g. based on correlation coefficient, will be performed.

## Section 4: Conclusions

Simulation of the 2-strip J-PET detector was performed using GATE software. The map of efficiency was determined using obtained results and compared with the map of geometrical acceptance.

The efficiency of our detector depends on many geometrical and physical factors. Simulations presented in this article indicate that the map of the total detection efficiency is strongly correlated with the detector geometry. More detailed study of the influence of each of the physical processes is underway.

The GATE software and its configuration on CIŚ cluster (e.g. configuration of output obtained with GATE software) must be optimized for shortening the time of performing the simulations.

# Acknowledgements

We acknowledge technical and administrative support by M. Adamczyk, T. Gucwa-Ryś, A. Heczko, M. Kajetanowicz, G. Konopka-Cupiał, J. Majewski, W. Migdał, A. Misiak, and the financial support by the Polish National Center for Development and Research through grant INNOTECH-K1/IN1/64/159174/NCBR/12, the Foundation for Polish Science through MPD programme, the EU and MSHE Grant No. POIG.02.03.00-161 00-013/09, Doctus – the Lesser Poland PhD Scholarship Fund and Marian Smoluchowski Kraków Research Consortium "Matter-Energy-Future".